\newcommand{\beq}{\begin{equation}}
\newcommand{\eeq}{\end{equation}}

\documentclass[aps,preprint,superscriptaddress,showpacs]{revtex4}
\usepackage{epsfig}
\usepackage{natbib}
\usepackage{amsmath}
\usepackage{times}
\usepackage{psfrag}

\usepackage{graphicx}
\usepackage{subfigure}
\usepackage{float}
\begin{document}
\title{Detection of noise-corrupted sinusoidal signals with Josephson junctions}

\author{Giovanni Filatrella}
\affiliation{CNR/SPIN and Dept. of Biological and Environmental Sciences, 
University of Sannio, Via Port'Arsa, 11, I-82100 Benevento, IT}
\author{Vincenzo Pierro}
\affiliation{Dept. of Engineering, University of Sannio, Corso Garibaldi, 107, I-82100 Benevento, IT}

\date{\today}

\begin{abstract}
We investigate  the possibility of exploiting the speed and low noise features of Josephson junctions for 
detecting sinusoidal signals masked by Gaussian noise. 
We show that the escape time from the static locked state of a Josephson junction is very sensitive to 
a small periodic signal embedded in the noise, and therefore the analysis of the escape times can be employed to reveal 
the presence of the sinusoidal component. We propose and characterize two detection strategies: in the first the initial phase is 
supposedly unknown (incoherent strategy),
while in the second  the signal phase remains unknown but is fixed (coherent strategy).
Our proposals are both suboptimal, with the linear filter being the optimal detection strategy, 
but they present some remarkable features, such as resonant activation, that make  detection through Josephson junctions appealing
in some special cases. 
\end{abstract}

\maketitle

\newpage

\section{Introduction}

Detection of sinusoidal signals corrupted by Gaussian noise is, in principle, a completely solved problem. 
It can be proven that in the framework of statistical decision theory (Neyman-Pearson scheme) an adapted linear 
filter implemented via a Fourier transform is the optimal choice \cite{Helstrom}. Unfortunately it is not always possible to apply 
such optimal detection strategy, due to a huge amount of data to process or 
to the extreme weakness of the signal to be detect.
As an example we mention the all-sky all-frequency search for gravitational waves emitted by a pulsar \cite{Krolak}, the search
for continuous monochromatic signal in radio astronomy \cite{astronomy} and the detection of THz radiation \cite{Mittleman}. 
When the optimal choice is not applicable, suboptimal strategies should be compared \cite{Schutz}:
for  example  it has been proposed to use generic bistable systems \cite{Inchiosa,Gammaitoni}.
The general idea is the following: 
the oscillatory signal corrupted by noise is applied to a nonlinear bistable element that transforms the original 
signal to a new one, simpler to analyze.
In other words, instead of applying a linear matched filter to the source signal, one inserts a nonlinear element; such insertion cannot improve the 
overall performance  \cite{Pinto}, but can make detection 
easier or can reduce the computational or memory burden.

Among nonlinear elements (like generic bistable devices \cite{Inchiosa} and superconducting systems \cite{SIS}) a 
good candidate is the Josephson Junction \cite{Barone} (henceforth JJ) for two important reasons: 1) JJs are extremely fast elements, 
that can easily operate above 100GHz and up to the THz region \cite{Ozyuzer}; 2) JJs , being superconductive elements, 
can be cooled close to absolute zero or to the quantum noise limit \cite{Cosmelli}, thus lessening the additional 
contribution of thermal noise from the bistable element. 
The detection of small signals in the presence of noise with JJs
has been reported in two configurations: investigating the
 magnetic flux trapped by a JJ closed in a superconducting loop \cite{Glukhov,Cirillo} 
(thus forming a SQUID \cite{Barone}) or analyzing the switching from the metastable locked solution 
to the running state of an underdamped isolated JJ \cite{Sun}. 
The SQUID exhibits a standard double well potential equivalent to the classical case, 
well known in the context of stochastic resonance \cite{Gammaitoni89,McNamara} as a relevant phenomenon for signal detection \cite{Inchiosa,Rouse}, 
although suboptimal \cite{Pinto}.
However SQUIDs, even if they are very sensitive and well studied systems, meet a principal 
drawback in their limited bandwidth, typically bounded to few KHz 
\cite{Rouse,Hibbs}. The escape from the static solution of single JJ \cite{Sun}, 
being extremely fast, seems to be promising for signal processing. 
In this paper we aim to characterize such a device as a detector.  

The detection scheme that we propose is closely related to the general problem of passage over a time-dependent barrier.
More precisely, when the JJ passes the maximum of the time-dependent potential, switching from the static to the running solution,
it sets the first passage time across an absorbing barrier \cite{Risken}. 
So the mean time between the initial state and the switch (or absorption) can be interpreted as a mean first passage time. 
Signal detection is then based on the possibility of ascertaining if the first passage times are due to a 
constant barrier (no signal present) or by an oscillating barrier (signal is present).

\section{Model and detection strategies}

The tunnel of Cooper pairs and normal electrons  through a JJ biased with a 
sinusoidal signal of amplitude $S_0$, a noise signal $S(t)$ and an additive thermal noise $N(t)$ can be modeled by 
the Langevin equation \cite{Barone}:

\beq
C\frac{d^2\varphi}{dt^2}+\frac{1}{R}\frac{d\varphi}{dt}+I_c \sin(\varphi)=I_B+ S_0 \sin(\Omega t+\varphi_0)+
S(t)+N(t).
\label{eq:joseqphu}
\eeq

\noindent
Eq. (\ref{eq:joseqphu}) is called underdamped because of the presence of the second derivative, or 
non negligible capacitance. In normalized units the equation reads:

\beq
\ddot{\varphi}+
\alpha \dot{\varphi}+ \sin(\varphi)=\gamma+ \varepsilon \sin(\omega {\tilde t}+\varphi_0)+
\xi({\tilde t})+\eta({\tilde t}).
\label{eq:joseq}
\eeq

Here $\alpha=(\hbar/2eI_c C)^{1/2}/R$ is the dissipation ($I_c$, $R$ and $C$ are the critical current, the resistance
and the capacitance of the junction), $\gamma=I_B/I_c$, the dc bias due to the physical bias current $I_B$.
These parameters can be adjusted in the experiments, although it is not easy to control dissipation (determined by the normal resistance $R$) 
one could in principle insert shunt resistors to achieve the desired level of dissipation. 
We have set $\alpha =0.2$ and $\gamma = 0.6$ without performing any parameter optimization to 
devise the best compromise between signal detection performances and experimental simplicity: we have just fixed
 the parameters for numerical convenience.
Furthermore $\varepsilon=S_0/I_c$ is the amplitude of the ac signal,
$\omega_j=[2eI_c/(\hbar C)]^{1/2}$ the Josephson frequency, and $\tilde{t}=\omega_j t$  the normalized time. 
Overdot denote the derivative respect to $\tilde{t}$. 

In Eq. (\ref{eq:joseq}) two random terms appear,
$\eta$ that represents thermal current 
fluctuations of intensity $<\eta(\tilde{t})\eta(\tilde{t}')>=2\alpha \theta\delta(\tilde{t}-\tilde{t}')$,
($\theta=2e k_B T/(\hbar I_c)$ is the normalized temperature) and 
$\xi$ that represents an additive noise with autocorrelation function of
intensity $D$, viz $<\xi(\tilde{t})\xi(\tilde{t}')>=2D\delta(\tilde{t}-\tilde{t}')$, corrupting the external signal. 
To assume that the signal is only corrupted by an additive term is a simplification: 
noise can affect the signal in several ways, for instance as a bandpass noise, multiplicative noise,  phase noise, or as frequency fluctuations. 
We have focused on the simple case of additive noise (that is a paradigm in signal processing);
however we expect that the results can be indicative of the behavior also for other noise sources.
For instance frequency fluctuations of the drive in a
JJ can be treated, in some limits, as an additive noise \cite{FMP}. Quantum fluctuation contribute as an equivalent thermal source of temperature $\theta^*=e \omega_J/( \pi I_c)$ \cite{Affleck,Cirillo}, 
that can be confused with the stochastic effects \cite{devoret85}.
Thus it is in principle always possible to decrease the temperature, proportional 
to $\theta$, to reduce fluctuations to an unavoidable quantum noise level. 

A washboard potential is associated with Eq. (\ref{eq:joseq}) \cite{Barone,Benjacob}:

\beq
U(\varphi)=-\gamma \varphi - \cos(\varphi).
\label{washb}
\eeq

\noindent 
The washboard potential  (\ref{washb})
for $\gamma<1$ gives rise to a barrier \cite{Benjacob}:
\beq
\Delta U(\gamma)=2[\sqrt{1-\gamma^2} - \gamma \cos^{-1}(\gamma)].
\label{eq:barrier}
\eeq
If the oscillating current is zero ($\varepsilon=0$) for low noise
 ($\theta << D$, $D<<\Delta U$) escape occurs at a rate $r_0$ \cite{Risken} 
\beq
r_0 \propto \tau_K^{-1} \exp(\frac{\Delta U}{D}).
\label{kramer}
\eeq

\noindent Such a rate is related to the average escape time $\tau_0=1/r_0$ (
$\tau_K$ is the Kramer prefactor, see \cite{Risken}). The escape times can be directly measured in experiments \cite{devoret85, Zhang}. 
In fact when an underdamped  junction 
leaves the metastable zero voltage state it switches to a running state with which is associated a voltage. 
It is therefore possible to measure the time elapsed from the application of the signal to the escape. 

The possibility of experimentally determining the probability distribution of the 
escape times and the average escape rate is the key feature that we want to exploit 
for signal detection. More precisely, we propose to take advantage of the exponential dependence of 
the mean escape time upon the barrier height. The escape waiting time is highly sensitive to the 
amplitude of the signal \cite{Yu}  as noticed since the pioneering experiments \cite{devoret87} that
 reported the striking changes into the escape time distributions for the resonant activation of JJ.
For the case of no signal, the expected residence time distribution
is exponential, while for oscillating barriers the distribution is modified and the average changed. 
The phenomenon has been thoroughly analyzed for overdamped systems \cite{Guentz} with colored noise 
\cite{Augello} and  measured for underdamped systems \cite{Sun}.
 For square wave signals, where a sudden switch of the barrier (\ref{eq:barrier}) between two values occurs,
 a remarkably accurate estimate has been proposed \cite{Yu}. Unfortunately, when a sinusoidal signal 
is embedded in noise and the barrier (\ref{eq:barrier}) is modulated,
the formula of Ref. \cite{Yu} is no longer valid.
Moreover, the method used in Refs. \cite{Sun,Yu} assumes that the signal is applied with a known 
phase ($\varphi_0$ in Eq. (\ref{eq:joseq}) ) and therefore the 
escape time distribution is a function of the initial phase \cite{Sun}. 
In the context of signal detection, the frequency of the signal might be unknown, and therefore such approach could be inapplicable.
In this paper we propose to use two  procedures. The first can be implemented without knowledge of the phase ({\em incoherent detection}), 
and only relies on detection of the escape time and subsequent reset of the system to the bottom of the potential well (see Fig. 1). 
The second strategy ({\em coherent detection}) uses the phase parameter (that is supposed constant, although unknown) 
and hence is more similar to the approaches \cite{Sun,Yu}.

The procedure that we propose for the incoherent detection strategy is the following:
 a) the signal-noise is applied with an  unknown initial phase $\varphi_0$ in Eq. (\ref{eq:joseq}); 
b) when an escape occurs in the evolution of Eq. (\ref{eq:joseq})  (the junctions reaches the metastable state $U_{max}$) the time $\tau_i$
 necessary for such an escape is recorded and the signal-noise is arrested or disconnected from the system; 
c) the JJ is reset to the bottom of the potential well $U_{min}$; d) the signal-noise is reapplied, {\em i.e.}
 Eq. (\ref{eq:joseq}) is integrated adding $\omega \tau_i$  to the previous initial phase. 
We have numerically checked that after few iterations the memory of the initial phase $\varphi_0$ is lost 
and that the long term distribution of the escape times is independent of the choice of the initial phase.
It is therefore also possible to apply the incoherent strategy  if the initial phase is unknown.

In the second case (coherent detection strategy) one assumes  that the frequency $\omega$ of the signal is fixed. 
Under such an hypothesis, we adopt the following procedure (for square signals it corresponds to the procedure of Ref. \cite{Sun}): 
a) the signal-noise is applied with an  unknown initial phase $\varphi_0$ to a JJ that is at rest in the bottom of the potential well (\ref{washb}). 
The phase between the signal and the JJ  is therefore fixed, but concealed; 
b) when the escape occurs, the escape time $\tau_i$ to reach the metastable state $U_{max}$ is recorded and the signal-noise
 is only arrested after a time that is a multiple of the period $2\pi/\omega$ of the external radiation 
 to guarantee that the same (unknown) initial phase $\varphi_0$ of the signal is retrieved;
c) the JJ is reset to the bottom of the potential well $U_{min}$  to reproduce the same initial conditions as at point a); 
d) the signal is restarted, {\em i.e.} Eq. (\ref{eq:joseq}) is 
 reapplied, and it will have with respect to the JJ the same initial phase $\varphi_0$. 
We remark that the phase difference between the JJ and the signal, $\varphi_0$, is frozen but unknown. 
To interpret the data it is necessary to reproduce the results for all values of $\varphi_0$. In fact  
the distribution of the escape times depends upon the initial phase $\varphi_0$, as will be discussed in Sect. IV.

We remark on the main difference between the two strategies: when the system switches
from the locked to the running state in the incoherent strategy we  immediately stop the application of the noise-signal, while in the coherent strategy we  let the signal run to 
retrieve the initial phase. 

For both strategies  if the signal is digitally recorded, as it is the case for the all-sky-all-frequency search of 
gravitational waves emitted by a pulsar, a preliminary digital/analog conversion could be required to obtain a signal in 
the JJ frequency range. In this case it is possible to apply the recorded signal with a much faster time, 
in the GHz range or as fast as the electronics allows (while the real time signal might be slower) thus achieving a considerable speed up. 

Typical Complementary Cumulative Distribution Function (CCDF) of the escape times in presence of a sinusoidal signal 
($\varepsilon=0.2$, broken lines) and without signal ($\varepsilon=0$, solid line) 
have been simulated by numerical integration of Eq. (\ref{eq:joseq}) with the Euler method \cite{mannella} and are shown in 
Fig. \ref{CCDF}. 
Such a Figure can also be interpreted as the CCDF for the passage times of a particle in a well over an oscillating barrier (broken lines) or over a constant barrier (solid line). 

From the data it is clear that the CCDF for the two strategies, coherent and incoherent, are very similar, with just a change in the slope and therefore a change in the average escape time.

\section{Results for incoherent detection}

\begin{figure}
\centerline{\includegraphics[scale=0.3]{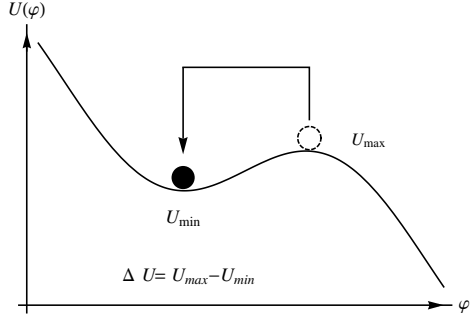}}
\caption{A picture of the potential barrier and of the detection procedure. 
When the phase (depicted as a dashed circle in figure) reaches the top of the barrier from the left,
the system is restarted with a suitable initial phase from the bottom (black disk in figure) of the 
potential well $U(\varphi)$. }
\label{model}
\end{figure}

\begin{figure}
\centerline{\includegraphics [keepaspectratio,width=8cm]{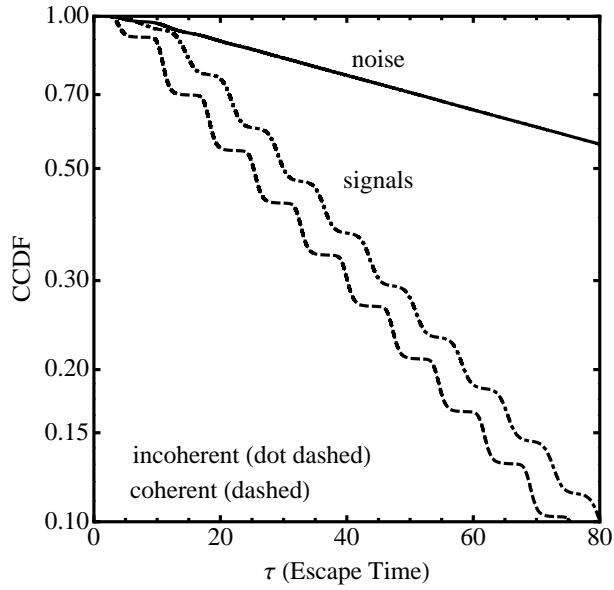}}
  \caption{The Complementary Cumulative Distribution Function (CCDF)
   of the escape time with signal $\varepsilon=0.2$ in the incoherent (dot-dashed line) or coherent ( dashed line) case. We also show the response to pure noise (solid line).The initial phase for the  coherent strategy is $\varphi_0=0$.
Parameters are: $\gamma=0.5$, $\alpha=0.2$, $\omega=0.86$, $D=0.05$, $\omega_0=(1-\gamma^2)^{1/4}\simeq 0.93$}
\label{CCDF}
\end{figure}

\begin{figure}
\begin{center}
\includegraphics[keepaspectratio,width=8cm]{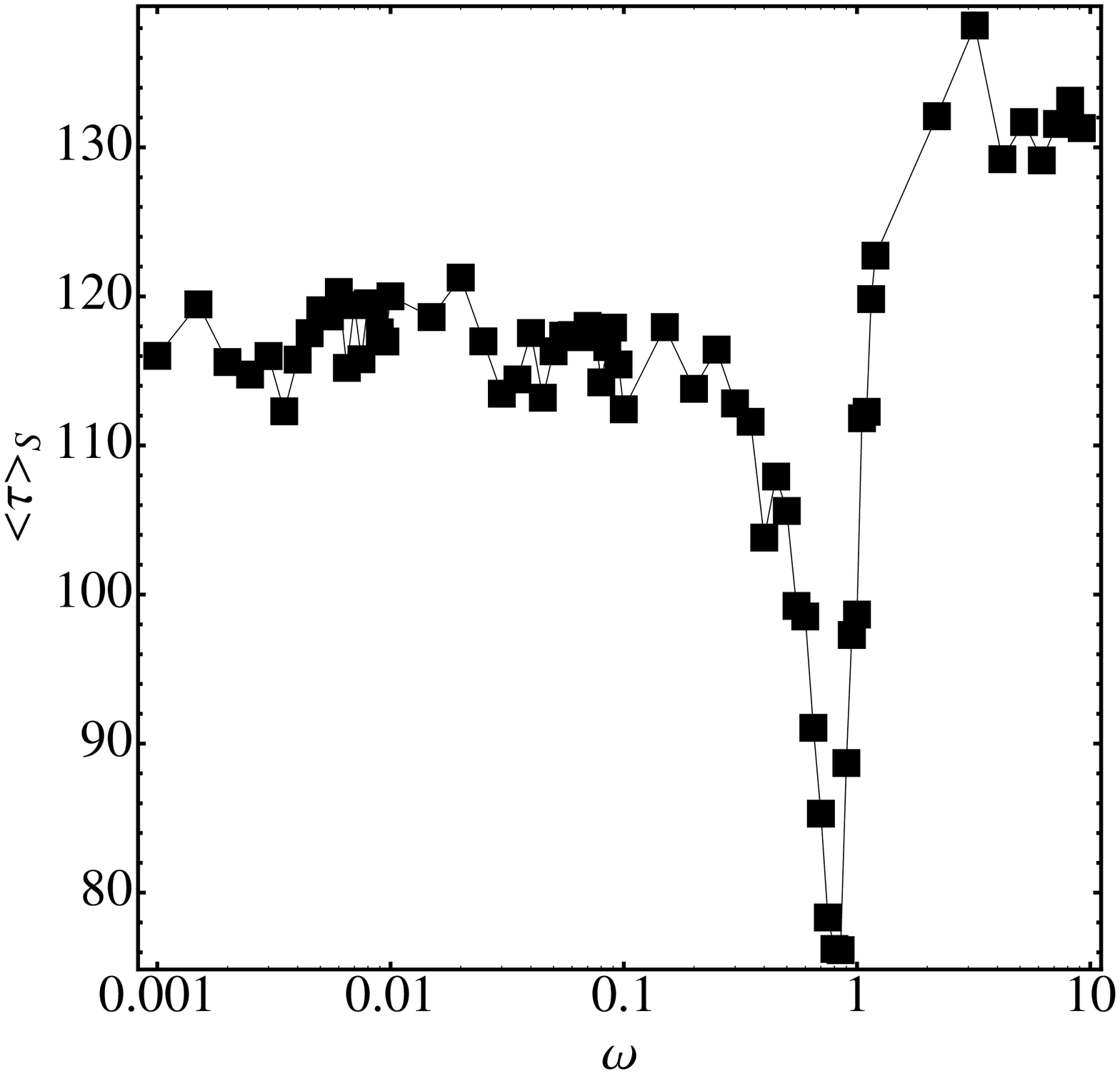}
\end{center}
\caption{ Typical estimated $\langle \tau \rangle_S$ as a function of $\omega$ ($T=2~10^5~2\pi/\omega_0$) for the incoherent detection strategy.
Parameters are: $\gamma=0.5$, $\alpha=0.2$, $D=0.05$, $\varepsilon=0.1$.
The average escape time in absence of the signal is $\tau_0=134.6$.}
\label{avetauuncoher}
\end{figure}

In Fig. \ref{avetauuncoher} we show the estimated average escape times through $U_{max}$ (see Fig. \ref{model}) vs $\omega$ of a phase
particle subject to an external drive and detected with the incoherent strategy (dot-dashed line of Fig. \ref{CCDF}). 
Around the normalized Josephson resonant frequency $\omega_0=(1-\gamma^2)^{1/4}$ the escape time is very sensitive to the external signal, 
in fact it exhibits a large dip, i.e. a pronounced deviation of the average escape time respect to the unperturbed value  
even for small $\varepsilon/\sqrt{D}$. (We recall that $\varepsilon/\sqrt{D}$ is proportional to the signal to noise ratio, SNR).
At low frequencies the deviations are less pronounced but still relevant, while for high frequencies 
one recovers the Kramer escape time of the unperturbed system \cite{Risken}. It is clear that there is a wide range of 
frequencies $(0,\omega_0)$ where the analysis of the estimated average escape times $<\tau>_S$ can give a clue to 
the presence of an external
 drive corrupted by noise, as they are different from the average of the escape times without the signal, $\tau_0$. 
 
\begin{figure}
\begin{center}
\includegraphics[keepaspectratio,width=8cm]{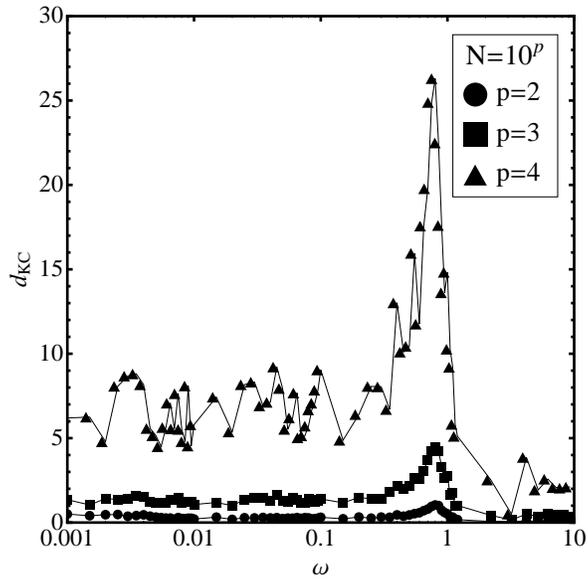}
\end{center}
\caption{
The dependence of the K-C index as a function of angular frequency for parameters $\gamma=0.5$, $\alpha=0.2$, $D=0.05$, $\varepsilon=0.1$ in the case 
of the incoherent detection strategy.
The curves displayed are for different observation time $T=2 \pi N/\omega_0$, where $N=10^p$.  }
\label{KComegauncoher}
\end{figure}

To make such analysis more quantitative, in Fig. \ref{KComegauncoher} we have adopted the 
Kumar-Carroll (K-C) index $d_{KC}$ \cite{KC}: 

\beq
d_{KC}= \frac{\mid \langle\tau\rangle_S-\langle\tau\rangle_N\mid}{
\sqrt{ \frac{1}{2}\left( \sigma^2(\tau)_S+\sigma^2(\tau)_N \right)}} ,
\label{KC}
\eeq
\noindent
where $\langle\tau\rangle_S$,$\langle\tau\rangle_N$ 
are the estimated average escape time over a prescribed interval $T$, with and without signal, respectively. We have also denoted with  
$\sigma(\tau)_S$,$\sigma(\tau)_N$ the corresponding estimated standard deviations.
The K-C index (\ref{KC}) is related to the receiving operating characteristics (ROC) of the detector, 
and as a rule of thumb is well approximated by the ROC of a matched filter with signal to noise ratio equal to $d_{KC}$ \cite{KC}. In Fig. \ref{KComegauncoher} it is evident, as expected, that $d_{KC}$ grows by increasing the observation time $T$.
The statistical analysis of Fig. \ref{KComegauncoher} confirms the result of Fig. \ref{avetauuncoher}, {\it i.e.} the existence of a peak at the geometric resonance $\omega_0$ ($d_{KC}\simeq 25$ at $T \simeq 10^4(2\pi/\omega_0)$), 
and of an interesting region for lower frequencies ($d_{KC}\simeq 5$), 
while the method seems inapplicable at frequencies above $\omega_0$  ($d_{KC}\simeq 1$). 
From Figs. \ref{avetauuncoher} and \ref{KComegauncoher} it is also clear that there is no evidence of stochastic resonance \cite{McNamara} due to resonant activation,
or prominent signal detection at  matching between the external drive frequency and the noise induced unperturbed escape rate, as  reported for instance in Ref.\cite{Sun}.
In contrast to stochastic resonance the results of Figs. \ref{avetauuncoher} and  \ref{KComegauncoher} only display a noise independent resonance at $\omega_0$. 
This difference can be ascribed to the peculiar manner in which the external signal is applied. 
In the incoherent strategy the system looses memory of the phase of the signal at the passage through the absorbing barrier 
(see Sect. II). 
Instead in Ref.\cite{Sun} the system is reset after each switch -- such reset corresponds to the coherent detection to be analyzed in Sect. IV. 

\begin{figure}
\begin{center}
\includegraphics[keepaspectratio,width=8cm]{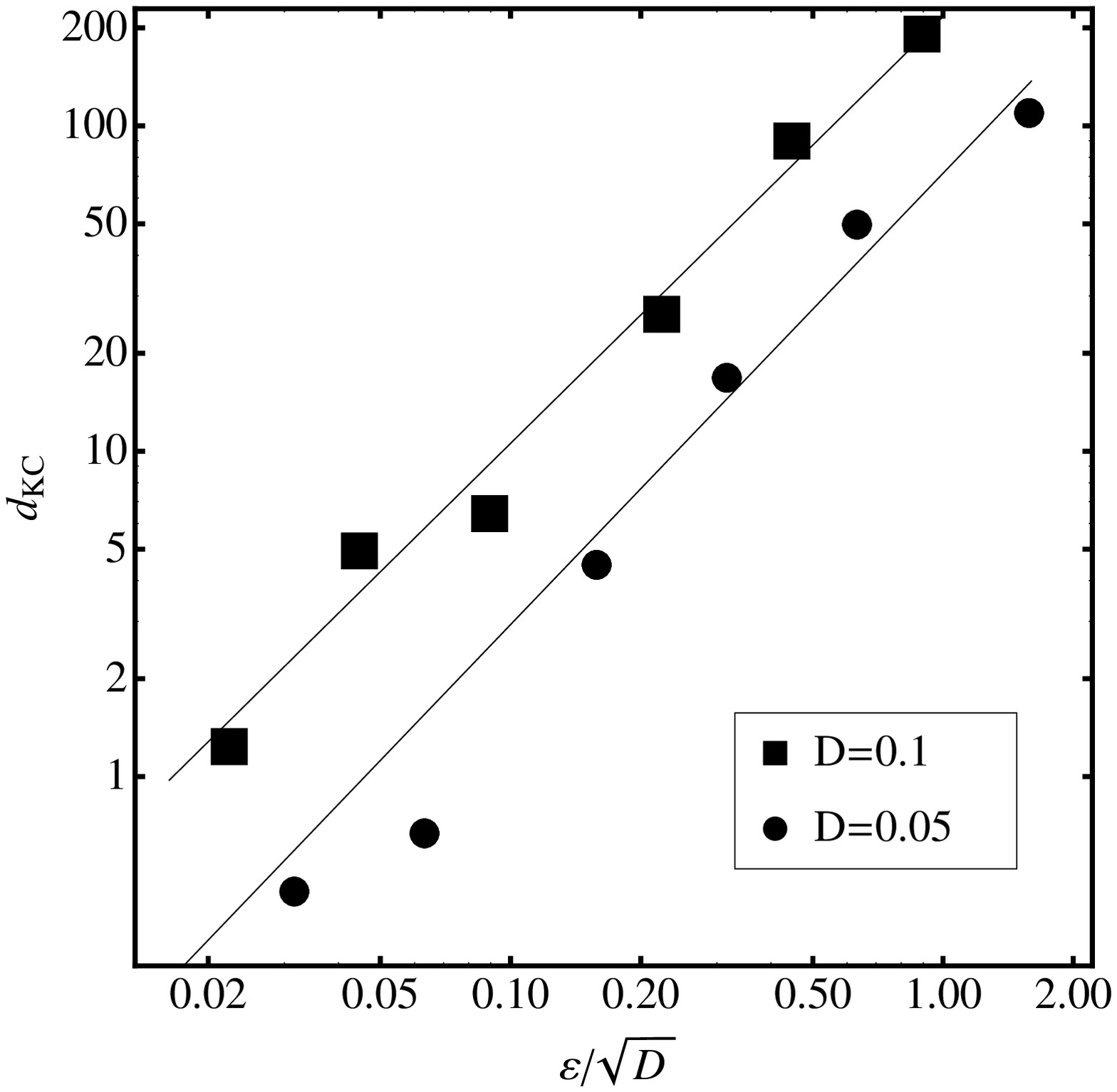}
\end{center}
\caption{
The dependence of K-C index as a function of scaled signal amplitude $\varepsilon/{\sqrt D}$ (proportional to the SNR)
 for  the incoherent detection strategy.The parameters used are
$\gamma=0.5$, $\alpha=0.2$, $\omega=\omega_0$, $T=2~10^5~2\pi/\omega_0$.}
\label{KCepsuncoher}
\end{figure}

In Fig. \ref{KCepsuncoher} we show the K-C index for several values of the ratio $\varepsilon/\sqrt{D}$. 
The purpose is to emphasize the behavior for small SNR, when reliable stochastic simulations are prohibitive. 
The data reduction demonstrates that the K-C index decreases roughly as an 
inverse power law of the SNR;
the best fit procedure gives an upper bound  of $\approx 1.5$ for the exponent. Extrapolating the results, it is thus possible to infer the behavior for SNR's
 lower than those reported in the figure. We recall that these performances 
refer to a specific signal duration $T=2~10^5~(2\pi/\omega_0)$ in Fig.\ref{KCepsuncoher}.
The K-C index increases extending the detection time $T$, insofar as the K-C index is roughly proportional to $\sqrt{T}$, see also Fig. \ref{KComegauncoher}.
Thus combining the numerical power estimated by Fig. \ref{KCepsuncoher} and the 
square root dependence of the K-C index captured by the data of Fig. \ref{KComegauncoher} it is possible to predict 
that to keep a fixed $d_{KC}$ while lowering $\varepsilon/\sqrt{D}$,
the detection interval should increase as $T \approx (\varepsilon/\sqrt{D})^{-3}$. 
Consequently, by extending the detection time $T$ one can also achieve the desired level of the K-C index for low values of SNR.
As expected, the matched filter requires shorter detection time ( $T \approx (\varepsilon/\sqrt{D})^{-2}$ ).

\section{Results for coherent detection}

\begin{figure}
\begin{center}
\includegraphics[keepaspectratio,width=8cm]{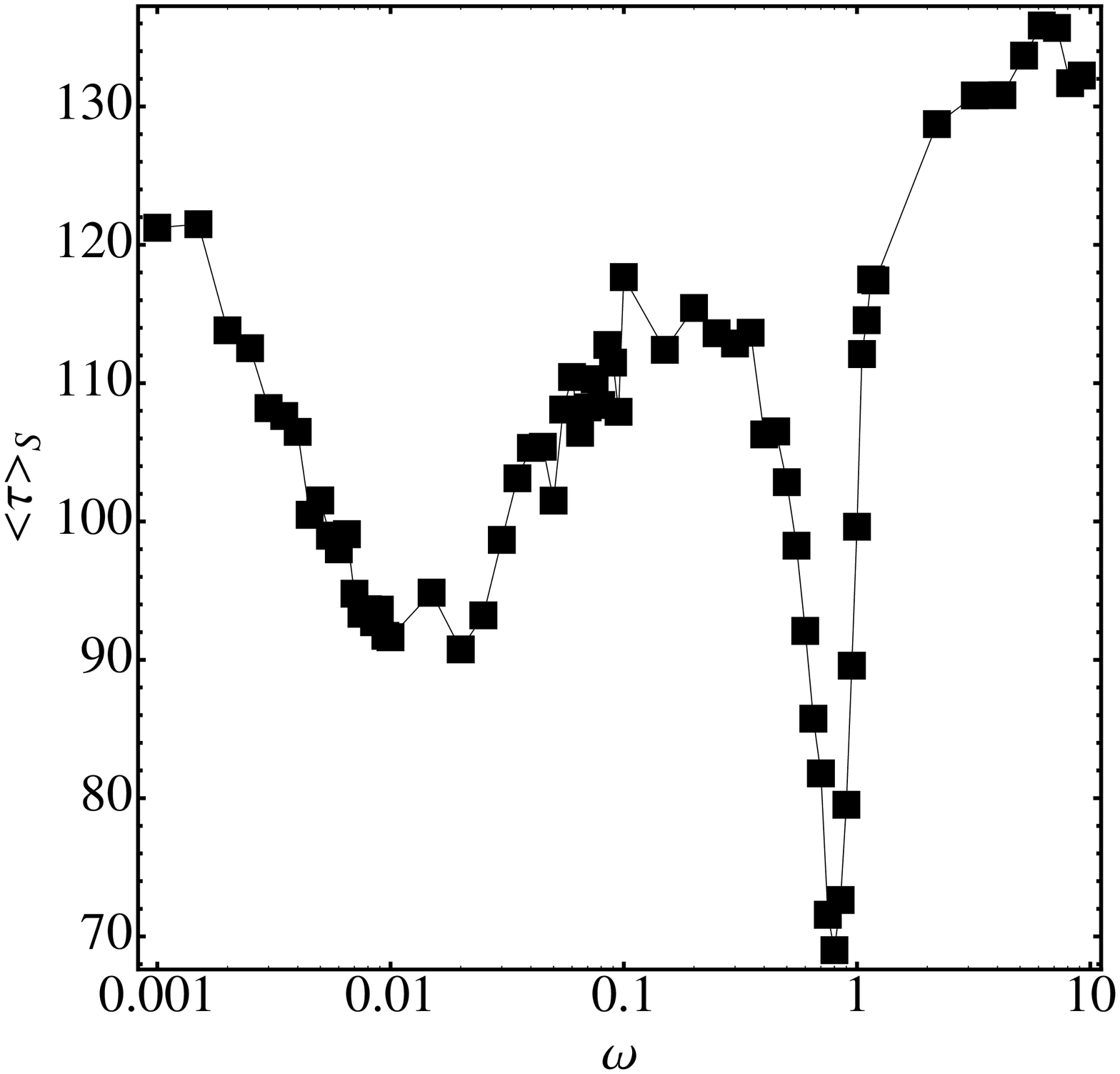}
\end{center}
\caption{ Typical estimated $\langle \tau \rangle_S$ as a function of $\omega$ ($T=2~10^5~2\pi/\omega_0$) for the coherent detection strategy.
Parameters are $\gamma=0.5$, $\alpha=0.2$, $D=0.05$, $\varepsilon=0.1$, $\varphi_0 = 0$.
The average escape time in absence of the signal is $\tau_0=134.6$.}
\label{avetaucoher}
\end{figure}

\begin{figure}
\begin{center}
\includegraphics[keepaspectratio,width=8cm]{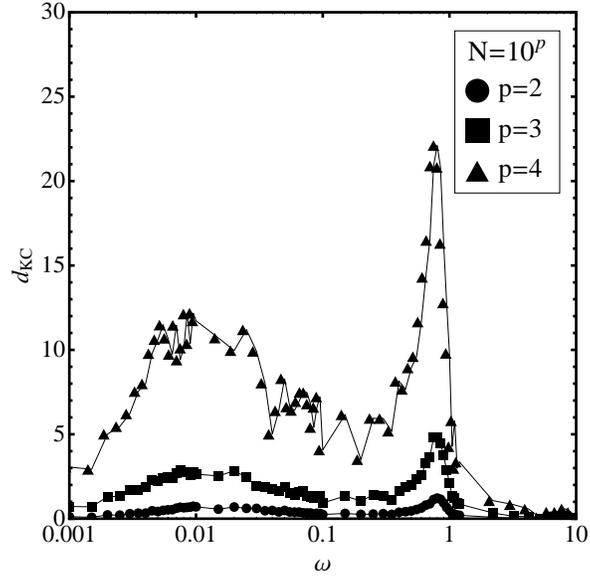}
\end{center}
\caption{
The dependence of K-C index as a function of scaled angular velocity for the coherent detection strategy. Parameters are $\gamma=0.5$, 
$\alpha=0.2$, $D=0.05$, $\varepsilon=0.1$, $\varphi_0=0$.
The curves displayed are for different observation time 
$T=2 \pi N/\omega_0$, where $N=10^p$.}
\label{KComegacoher}
\end{figure}

In Section II we discussed the possibility of detecting a signal at a predetermined frequency $\omega $ employing a coherent detection strategy.
 In such strategy the signal is always applied with the same initial phase.
The analysis of the estimated average escape time in presence of a signal with initial phase $\varphi_0=0$ is reported in Fig. \ref{avetaucoher}. The data show a remarkable dip at low frequency ($\omega \simeq 0.01$), that is not present in the incoherent detection strategy, see Fig. \ref{avetauuncoher}. To compare the performances we analyze the K-C index (\ref{KC}) in Fig. \ref{KComegacoher}. This analysis, as does the simpler analysis of Fig. \ref{avetaucoher}, confirms the existence of an interesting region ($d_{KC} \simeq 10$) at low frequency ($\omega \simeq 0.01$) that is not present in the incoherent strategy (see Fig.\ref{KComegauncoher}).

\begin{figure}
\begin{center}
\includegraphics[keepaspectratio,width=8cm]{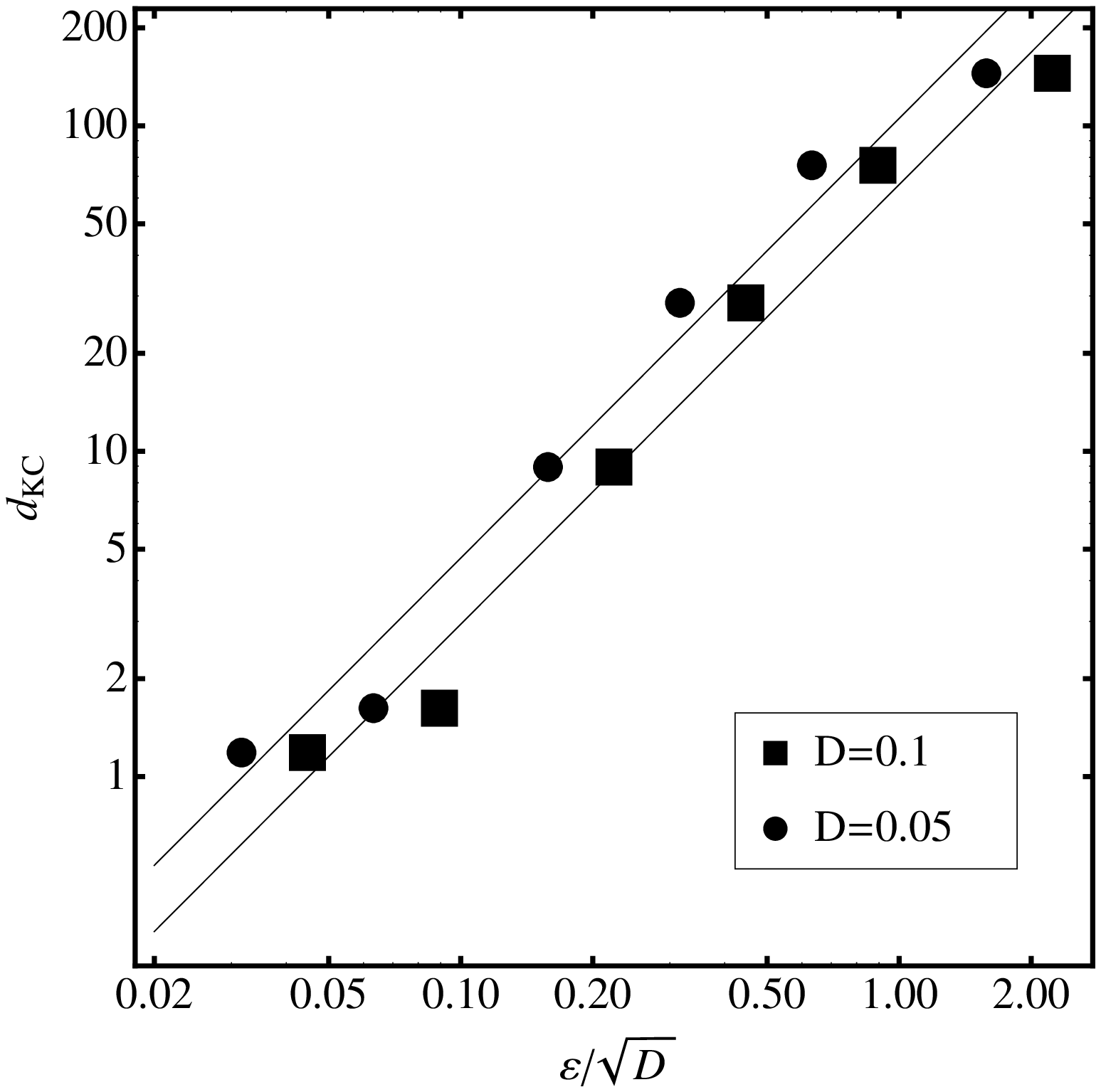}
\end{center}
\caption{
The dependence of K-C index as a function of scaled signal amplitude $\varepsilon/{\sqrt D}$  (proportional to the SNR) for the coherent detection strategy. The parameters used are $\gamma=0.5$, $\alpha=0.2$, $\omega=\omega_0$, $T=2~10^5~2\pi/\omega_0$. }
\label{KCepscoher}
\end{figure}

In Fig. \ref{KCepscoher} we show  the dependence of the K-C index as a function of the signal scaled 
amplitude $\varepsilon / \sqrt{D}$ (compare to  Fig. \ref{KCepsuncoher} for the incoherent analysis).
This analysis leads to the same conclusion as for the incoherent strategy, {\it i.e. } by increasing the detection time $T \approx (\varepsilon/\sqrt{D})^{-3}$ one can achieve the desired level of the K-C index.

\begin{figure}
\begin{center}
\includegraphics[keepaspectratio,width=8cm]{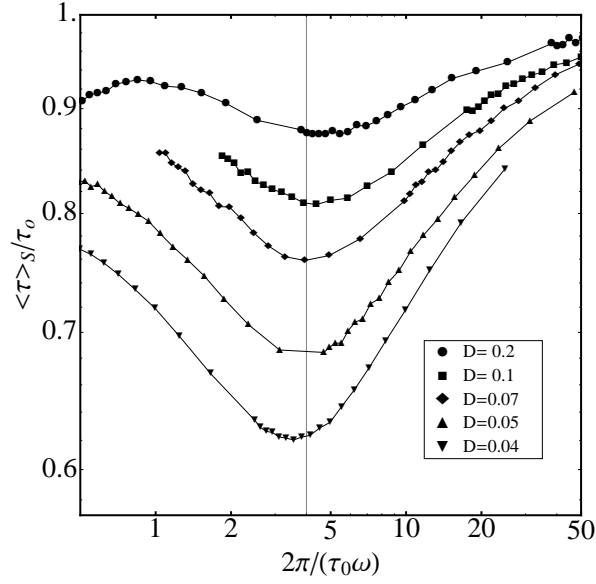}
\end{center}
\caption{
Scaled estimated average escape time  
$\langle \tau \rangle_S$ in presence of signal, vs scaled period $2 \pi/\tau_0 \omega$ for
different values of $D$ in the case of coherent strategy (initial phase $\varphi_0=0$).
The parameters used are $\gamma=0.5$, $\alpha=0.2$, $\varepsilon=0.1$,
$\tau_0$ is the $\varepsilon=0$  escape time. The vertical grid line indicates the resonant condition (\ref{matching}).}
\label{resovsnoise}
\end{figure}

\begin{figure}
\begin{center}
\includegraphics[keepaspectratio,width=10 cm]{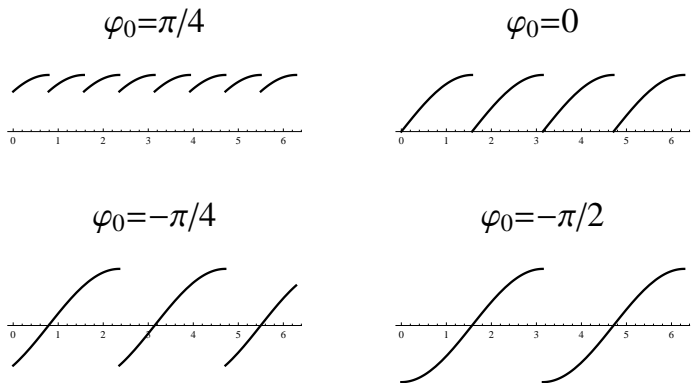}
\end{center}
\caption{
Effective waveforms for a system prepared with initial phase 
$\varphi_0=-\pi/2$,$-\pi/4$, $0$, $\pi/4$. The figures show the sinusoidal drive between the initial phase $\varphi_0$ and the maximum of the signal, $\varphi=\pi/2$.} 
\label{sintrunc}
\end{figure}

\begin{figure}
\begin{center}
\includegraphics[keepaspectratio,width=9 cm]{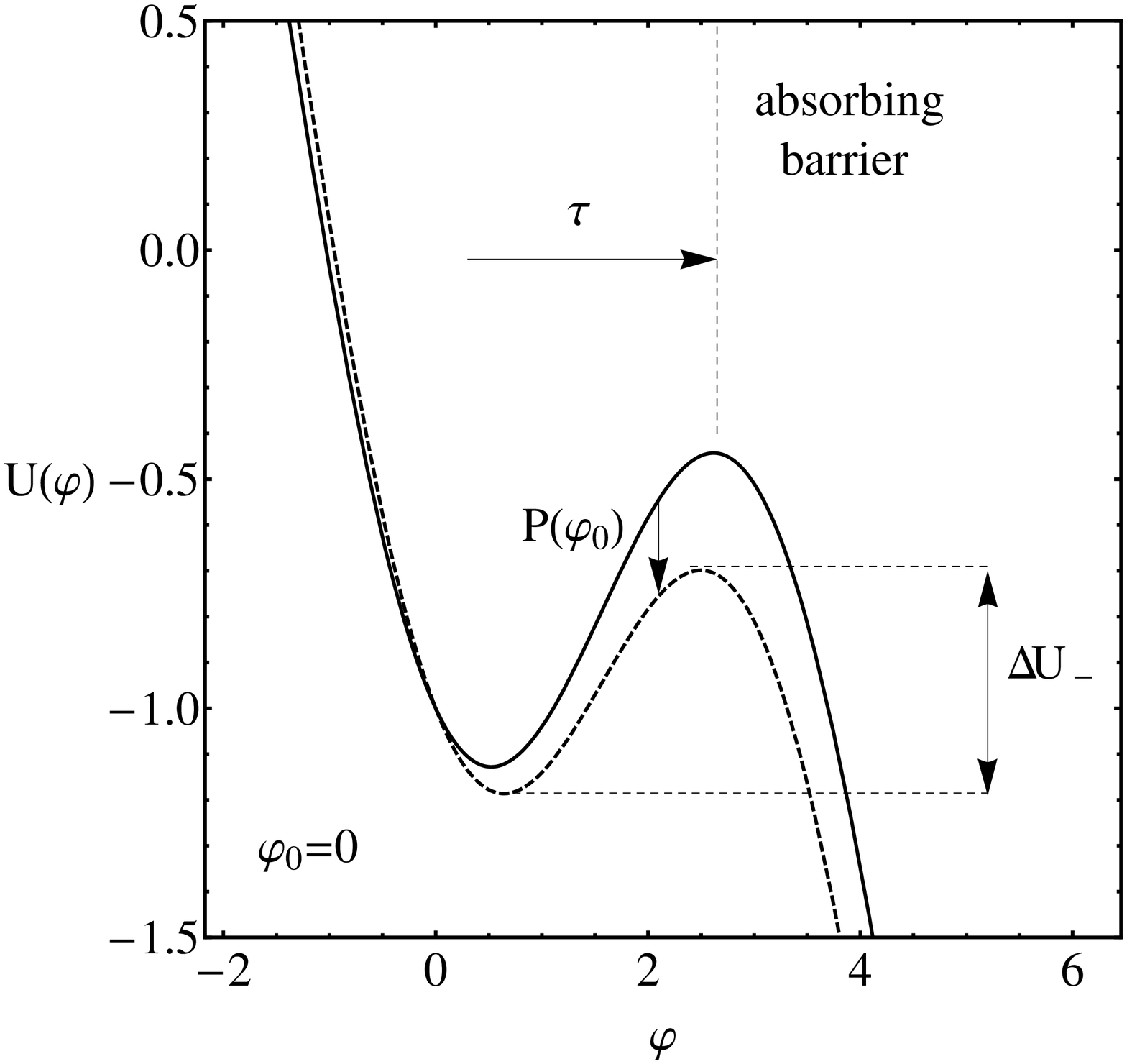}
\includegraphics[keepaspectratio,width=9 cm]{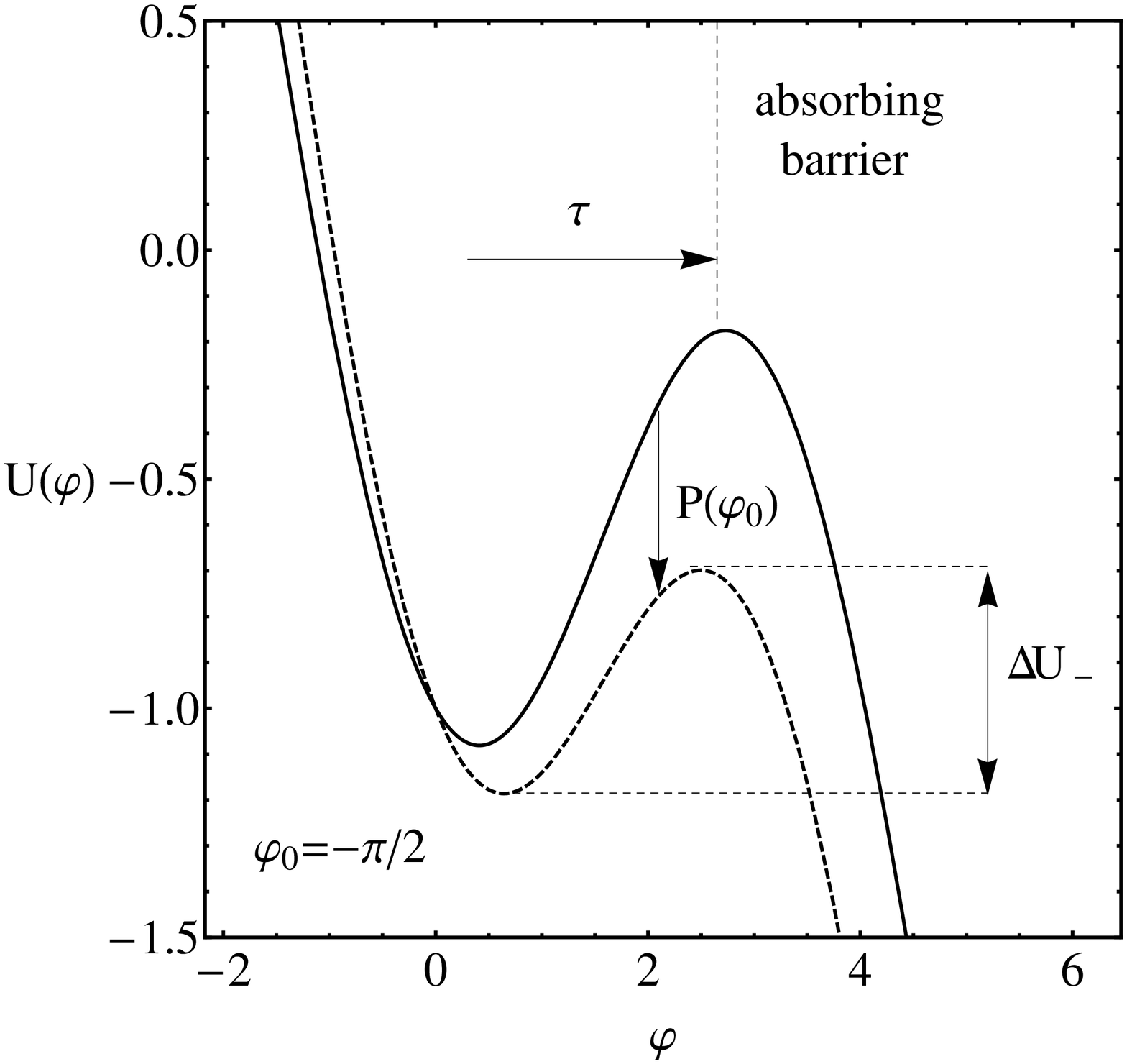}
\end{center}
\caption{The figures show the JJ potential $U(\varphi)=-\bar{\gamma} \varphi - \cos(\varphi)$ 
as a function of the phase. We consider $\bar{\gamma}=\gamma+\varepsilon \cos(\varphi_0)$ with initial phase $\varphi_0=0$ (top) or $\varphi_0-\pi/2$ (bottom);
for illustrative purpose $\gamma=0.5$ and $\varepsilon=0.1$.
The time $P(\varphi_0)$ is defined as the time between the application of the signal (with initial phase $\varphi_0=0,-\pi/2$)
 and the lowest height of the barrier, $\Delta U_-$. The first passage time $\tau$ is defined as the time to overcome the maximum of the barrier (the 
 vertical dashed line). 
 The resonance defined by Eq. (\ref{resP}) states that a minimum of the passage time occurs when are equal the time in which the external signal 
 reduces the barrier to the lowest value and the escape time from such  barrier, assumed to be static. 
}
\label{absbarr}
\end{figure}

\begin{figure}
\begin{center}
\includegraphics[keepaspectratio,width=8cm]{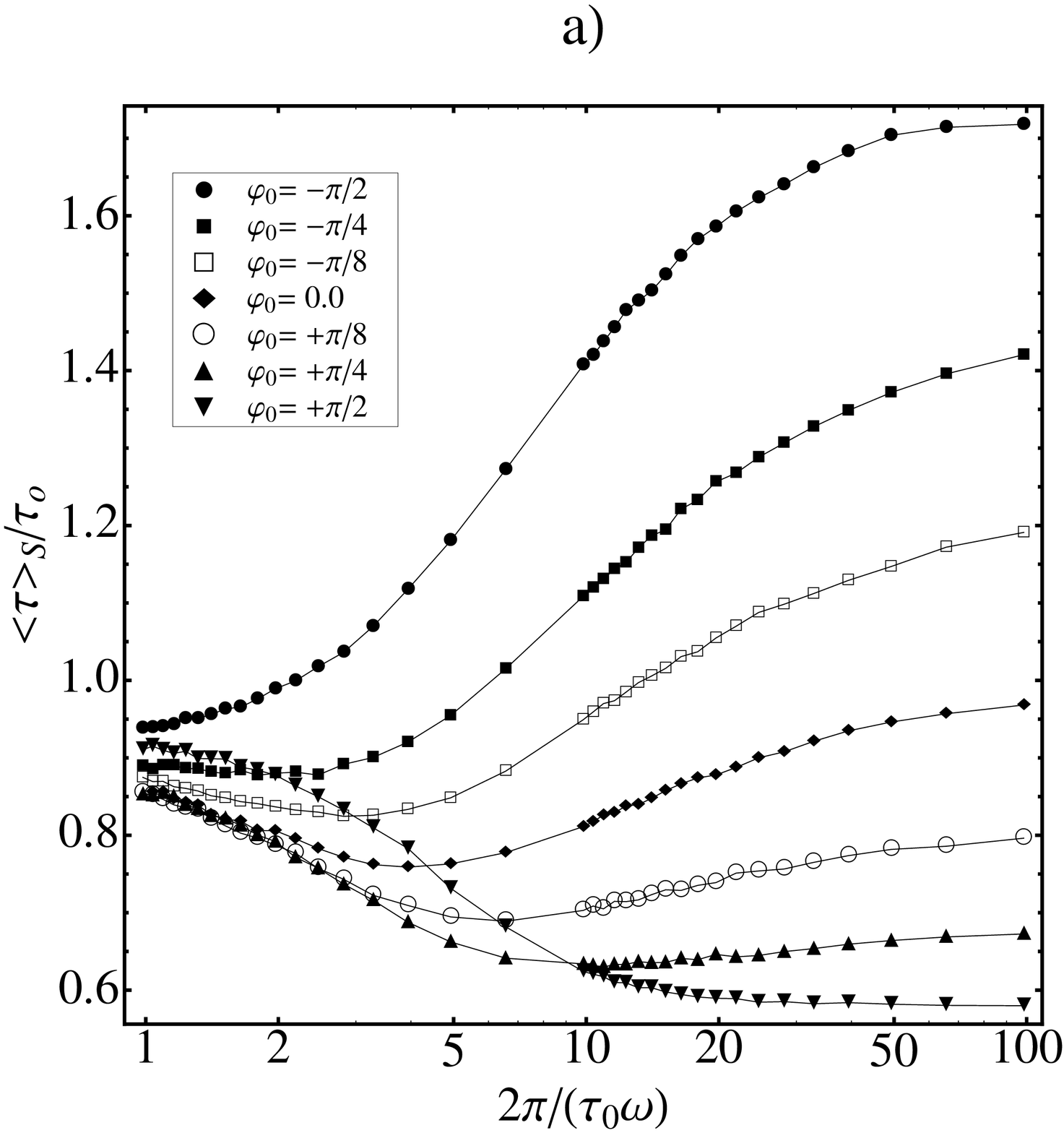}
\includegraphics[keepaspectratio,width=8cm]{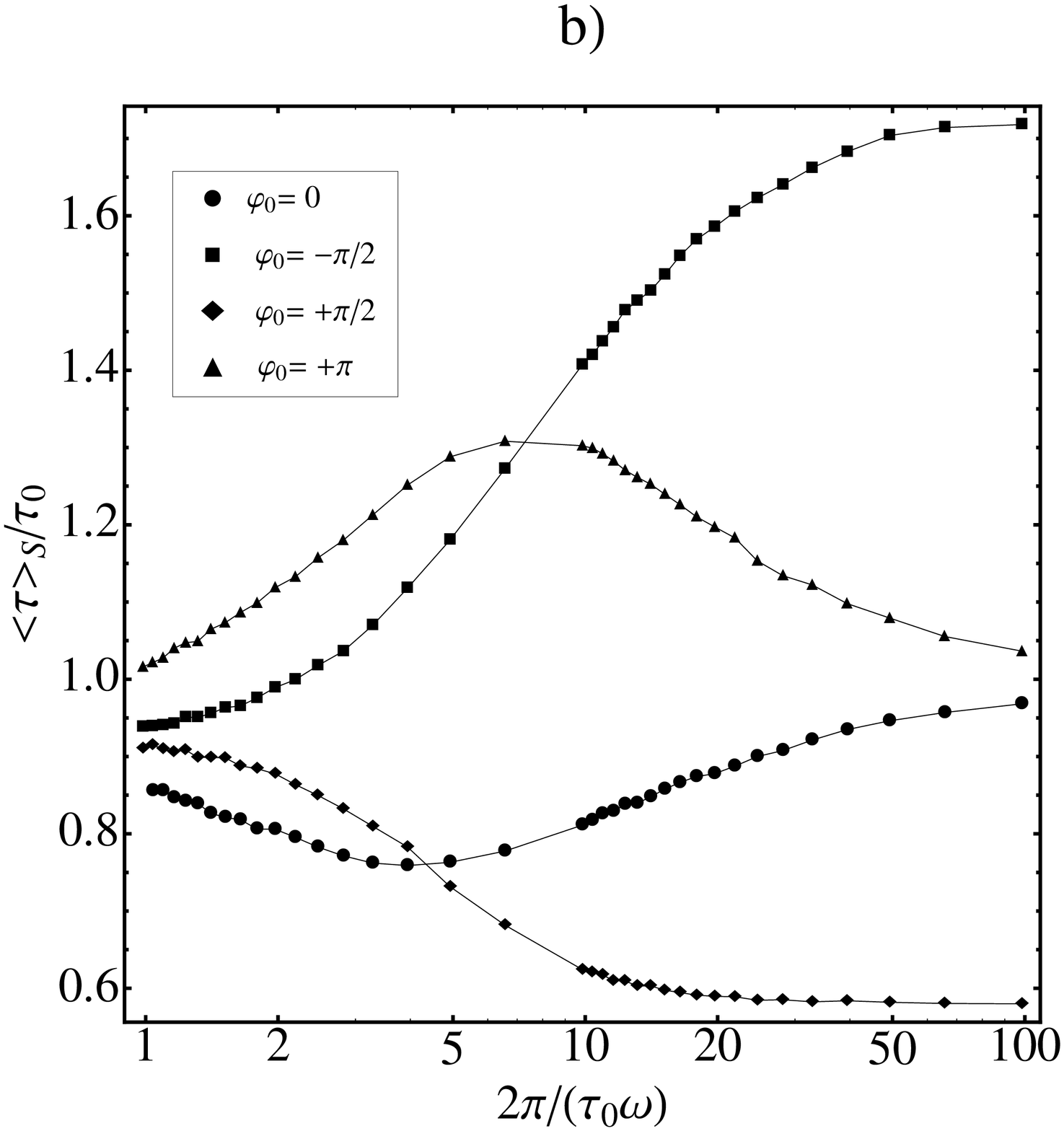}
\end{center}
\caption{
The dependence of the average escape time  $\langle \tau \rangle_S$  (scaled to $\tau_0$
) for the coherent detection strategy  as a function of signal scaled period  $2 \pi /\omega$ for different values of the initial phase $\varphi_0$. 
 In panel a) we show  the escape times when the effective waveform is monotonic, {\it i.e.} $-\pi/2 \le \varphi_0 \le \pi/2$; in panel b) we show the four limit cases $\varphi_0=0, \pi/2, \pi, -\pi/2$. We underline that for $\varphi_0 = \pi$ the non monotonic behavior of the effective waveform
( the maximum  is not reach  in a single ramp-up) leads to a maximum in the escape time.
The parameters used are $\gamma=0.5$, $\alpha=0.2$, $\varepsilon=0.1$, $D=0.07$.
}
\label{phase1}
\end{figure}

A second resonant condition for the coherent detection strategy occurs at the angular frequency:

\beq
\omega = C\left(\varphi_0\right) \left( \frac{2\pi}{\tau_0} \right),
\label{matching}
\eeq
where $\tau_0$ is the escape time of Eq. (\ref{eq:joseq}) for $\varepsilon=0$. 
\noindent The factor $C(\varphi_0)$ reads $1/4$ for $\varphi_0=0$ and is different from the frequency condition
 of the standard stochastic resonance \cite{Gammaitoni}. This difference has been also reported in the experimental findings of Ref. \cite{Sun,Yu}.
In Fig. \ref{resovsnoise} the frequency relation (\ref{matching}) is further
elucidated: we show for different values of $D$ the estimated escape time. From the figure 
it is clear that the largest deviation of the signal induced average escape time $<\tau>_S$ occurs at the matching condition described by 
Eq. (\ref{matching}) over a wide range of noise intensity, $0.04 \le D \le 0.2$. Fig. \ref{resovsnoise} thus indicates the presence of a stochastic
 resonance between the noise (which determines $\tau_0$) and the frequency $\omega$ of the external signal. 
We ascribe the change in Eq. (\ref{matching}) with respect to the traditional stochastic resonance to the peculiar choice of initial conditions. The reset of the initial conditions changes the effective waveform of the signal, see Fig. \ref{sintrunc}. 
Let us define $P(\varphi_0)$ as the time from the initial phase and the phase that corresponds to the maximum of the signal and therefore to the minimum of the barrier (\ref{eq:barrier}).
Assuming that the effective waveform of the signal is of the type depicted in Fig. \ref{sintrunc}, the matching condition for stochastic resonance shown in Fig. \ref{absbarr} is that the barrier reaches a minimum in a time $P(\varphi_0)$ close to the noise induced escape from the lowest energy well. Such resonant condition  depends upon the initial phase $\varphi_0$; in Fig. \ref{phase1} we report the scaled deviations of the average estimated escape time in presence of the signal as a function of the ratio $2\pi/\tau_0\omega$ for different values of the initial phase $\varphi_0$. We hypothesize that the horizontal axis positions of the relative minimal $C(\varphi_0)^{-1}$ in Fig. \ref{phase1}a correspond to the resonant condition of Fig.\ref{absbarr}. 
In other words, we observe that a resonant condition occurs when the time $\tau_0 (\Delta U_-)$ to escape the minimum barrier (denoted by $\Delta U_-$ in Fig.\ref{absbarr}) matches the time in which the oscillating barrier reaches a minimum: 

\beq
\tau_0(\Delta U_-) = P(\varphi_0).
\label{resP}
\eeq

\begin{figure}
\begin{center}
\includegraphics[keepaspectratio,width=8cm]{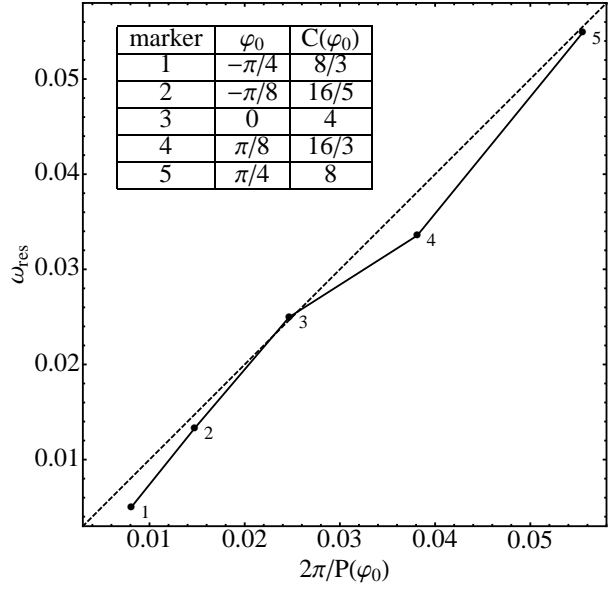}
\end{center}
\caption{
The dependence of the resonant frequency as a function of $2\pi/P(\varphi_0)$, proportional to the inverse  time between the initial phase and the maximum of the signal, see Fig. \ref{absbarr}. 
The dashed line corresponds to the prediction of Eq. (\ref{resP}). 
Parameters are $\gamma=0.5$, $\alpha=0.2$, $\varepsilon=0.1$, $D=0.07$.
}
\label{omegattiva}
\end{figure}

\noindent
The resonance phenomenon described by Eq. (\ref{resP}) is evident in Fig.\ref{phase1}a where the barrier reaches a minimum in
a single ramp-up. Eq.  (\ref{resP})  has been  compared with numerical data in Fig. \ref{omegattiva} and shows  excellent agreement.
This simple picture does not hold when the barrier non-monotonically reaches the minimum; in fact in Fig.\ref{phase1}b the
condition (\ref{resP}) is not valid for $\varphi_0=\pi$.

We wish to underline the following remarks about Eq. (\ref{resP}):
\begin{itemize}
\item[1)] The prediction of Eq. (\ref{resP}) in the case $\varphi_0=0$, $P(\varphi_0)=(1/4)(2\pi/\omega_0)$, correctly describes
the experimental finding of an harmonically driven underdamped JJ , see Ref. \cite{Yu};
\item[2)] When the barrier decreases monotonically ($-\pi/2 \le \varphi_0 \le \pi / 2 $) and the escape occurs in a single 
ramp-up of the signal (see Fig. \ref{sintrunc}), the physical situation is similar to a fluctuating barrier \cite{Yu,Boguna,Porra,Doering}:
the time in which the washboard potential reaches the minimum, $P(\varphi_0)$ in our notation,  plays the role of the barrier flipping rate in Refs. \cite{Yu,Boguna,Porra,Doering}. 
\item[3)] By varying the bias current $\gamma$ (which is an external parameter) one can set the most appropriate barrier (see Eq.s (\ref{eq:barrier}) and (\ref{kramer}) ) to match
the resonant condition (\ref{resP}).
\end{itemize}

A qualitative argument to understand the behavior when the signal does not reach a maximum in a single ramp-up (see Fig.\ref{phase1}b) can be sketched following 
Ref. \cite{Porra}, where it has been noticed that a resonant activation occurs in two cases: 
\begin{itemize}
\item[a)] The low and high frequency limits of the escape times are identical: an extremum occurs if  the derivative with respect to $\omega$ of the barrier is nonzero;
\item[b)] The low and high frequency limits of the escape times are different: an extremum occurs if the derivative with respect to $\omega$ of the escape time is opposite to the difference of the limits.
\end{itemize}

The low  frequency limit of the escape time
can be estimated with an  heuristic argument (see the Appendix for details ):
\beq
\tau(\varphi_0,\omega)  \approx
\tau_K
e^{-\frac{\Delta U(\varepsilon \sin
   (\varphi_0)+\gamma)}{D}} 
\left(1+
A\omega\varepsilon
\cos (\varphi_0) \right)
+O((\omega 
 \tau(\varphi_0,\omega) )^2),
\label{tauomegasmall}\eeq
\noindent
where  $\tau_K$ is the Kramer prefactor  and the  positive constant $A$ reads:
\beq
A=-
\frac{ \tau_K}{D} 
e^{-\frac{\Delta U(\varepsilon \sin
   (\varphi_0)+\gamma)}{D}} 
\Delta U'(\varepsilon \sin(\varphi_0)+\gamma).
\eeq

\noindent The high frequency limit (see the Appendix) is given by the time independent potential:

\beq
\tau(\varphi_0,\omega) \approx 
\tau_K
\exp 
\left[ -\frac{\Delta U(\gamma)}{D}
\right] + O\left(\frac{1}{\omega 
 \tau(\varphi_0,\omega) }\right).
\label{tauomegahigh}
\eeq

\noindent A consequence of Eqs. (\ref{tauomegasmall}) and (\ref{tauomegahigh}) is that for $0 \le \varphi \le \pi$ the low frequency limit of the 
barrier is lower than the high frequency limit, and therefore the escape time will be longer ( the reverse situation will be observed in the 
range $-\pi \le \varphi \le 0$ ). 
From  Eq. (\ref{tauomegasmall}) we are able to compute the following derivative
\beq
\frac{d\tau(\varphi_0,\omega) }{d\omega} \approx
\tau_K
e^{-\frac{\Delta U(\varepsilon \sin
   (\varphi_0)+\gamma)}{D}} A \varepsilon
\cos (\varphi_0) + O\left(\omega \tau(\varphi_0,\omega) \right).
\label{eq:derescape}
\eeq
Eq. (\ref{eq:derescape}), valid in the low $\omega$ regime,  shows that 
 the corrections to the zero frequency limit 
$\tau(\varphi_0,0)\approx \tau_K e^{-\frac{\Delta U(\varepsilon \sin
   (\varphi_0)+\gamma)}{D}} $ have the same sign of the
derivative of the external drive,  i.e. $\mbox{sign}(\cos\varphi_0)$. 
The results of Fig.\ref{phase1} confirm this picture: a resonance, or a non monotonic behavior
is observed for all initial phases but $\varphi_0 = \pm \pi/2$ (where the signal derivative vanishes and one should 
consider higher order corrections) and is most evident for  $\varphi_0 = 0,\pi$ - where the signal derivative is at a maximum -. 
In the most effective cases , $\varphi_0=0$ and $\varphi_0=\pi$, the former leads to the condition (\ref{resP}) with $C(\varphi_0)$
correctly predicted by Eq. (\ref{matching}) (see also Fig.\ref{omegattiva}), while the latter ($\varphi_0=\pi$) exhibits a different frequency
relation when the signal maximum not reached in a single ramp-up.

\section{Conclusions}

We employed the escape times of a JJ 
from the locked state as a statistic tool to detect sinusoidal signals
corrupted by noise. We have determined the main features of such a statistical
 detector for both incoherent and coherent detection. 
For the former  we  found that the detector: $i)$  is most sensitive to the signal
  in proximity to the junction resonance, 
$ii)$  shows a clearly asymmetric behavior in frequency: very low sensitivity above the resonance, 
and relatively good performance for lower frequencies. 
In the coherent case, which requires {\it a priori } knowledge  of the applied frequency, the performance
 depends upon the initial phase. For an appropriate choice of the initial phase it is also evident a resonant activation, {\it i.e.} a matching 
 condition between the noise induced escape time and the external frequency, Eq. (\ref{resP}),  well fits the numerical data.
Our analysis of the role of the initial phase extends the previously observed resonant escape \cite{Yu}. 
Future research could be directed towards:
$i)$ development of better detection strategies;
$ii)$ coupling two or more elements to exploit the properties of arrays of JJs \cite{2DJJ};
and $iii)$ a more accurate  analytical treatment of phase dependent resonant activation
phenomenon.

We remark that the detection scheme based on JJs is suboptimal, but could prove fast and capable of 
operation at very low temperatures, with low intrinsic noise. It is therefore in niche applications
where speed and reduced noise are essential that this approach could be considered. 
Finally, a few words of caution: while we have analyzed the physical principles underlying 
JJs as possible detectors, practical applications require a deeper analysis
of circuit design and of technological limitations.

We wish to thank S. Pagano and I. M. Pinto for fruitful discussions
and R. Newrock for a critical reading of the manuscript.
\newpage

\section*{Appendix}
In this Appendix we give the asymptotic behavior of $\tau(\varphi_0,\omega) $ in the slow and fast frequency
limit ($\omega \rightarrow 0$ and $\omega \rightarrow \infty$ ) . 
To compute the slow frequency behavior of the average escape time 
we use, following Ref. \cite{McNamara}, periodically modulated escape rates of Arrhenius type.
In this connection we have:
\beq
\tau(\varphi_0,\omega)  \approx
\langle \tau_K \exp[ -\frac{\Delta U(\tilde{\gamma}(\tilde{t}))}{D}] \rangle 
\eqno{(A1)}
\nonumber
\eeq
\noindent where $\tilde{\gamma}(\tilde{t})=\gamma +\varepsilon \sin(\omega \tilde{t}+\varphi_0)$
is the time dependent bias and $\tau_K$ is the Kramer escape rate prefactor. Since we are interested in computing
  the asymptotic limit $\omega \rightarrow 0$,
 there exists a frequency such that
$\omega \tilde{t} \ll \omega 4 \langle \tau \rangle \ll 1 $, where, due the exponential like
tail of the escape time distribution, the probability of occurrence of an escape time greater than $4 \langle \tau \rangle$ is negligible.
Accordingly, we expand the average escape time in Taylor series in the variable $\omega \tilde{t}$ 
and due to the exponential like distribution we approximation have the estimates $\langle \tilde{t} \rangle  \approx \tau(\varphi_0,\omega) $, 
$\langle \tilde{t}^2 \rangle \approx 2\tau(\varphi_0,\omega)^2$. The Taylor expansion of Eq. (A1) reads:
\beq
 \tau(\varphi_0,\omega) \approx
\tau_K
e^{-\frac{\Delta U(\varepsilon \sin
   (\varphi_0)+\gamma)}{D}} 
\left(1- 
\frac{\varepsilon \omega }{D} 
 \tau(\varphi_0,\omega)
\cos (\varphi_0) 
\Delta U'(\varepsilon \sin(\varphi_0)+\gamma)\right)
+O((\omega  \tau(\varphi_0,\omega))^2).
\eqno{(A2)}
\nonumber
\eeq 
In the previous equation $\tau(\varphi_0,\omega)$ appear on the right side of the formula $(A2)$,
it can be consistently eliminated by an iterative substitution (and truncation) of $(A2)$ in itself.
By defining the parameter :
\beq
A=-
\frac{ \tau_K}{D} 
e^{-\frac{\Delta U(\varepsilon \sin
   (\varphi_0)+\gamma)}{D}} 
\Delta U'(\varepsilon \sin(\varphi_0)+\gamma)
\eqno{(A3)}
\nonumber
\eeq
Eq. (A2) can therefore be written:
\beq
 \tau(\varphi_0,\omega) \approx
\tau_K
e^{-\frac{\Delta U(\varepsilon \sin
   (\varphi_0)+\gamma)}{D}} 
\left(1+
A\omega\varepsilon
\cos (\varphi_0) \right)
+O((\omega 
 \tau(\varphi_0,\omega) )^2).
\eqno{(A4)}
\nonumber
\eeq
By using (\ref{eq:barrier}), the condition $A>0$ holds.

The asymptotic limit $\omega \rightarrow \infty$ is obtained by noting that fast oscillation
can not be followed by the system dynamic \cite{Risken}. We have:
\beq
\tau(\varphi_0,\omega) \approx
\tau_K
 \exp\left[ -\frac{\Delta U(\gamma)}{D}\right] + O\left(\frac{1}{\omega 
\tau(\varphi_0,\omega) }\right).
\eqno{(A5)}
\nonumber
\eeq

\newpage


\newpage

\newpage

\newpage

\end{document}